\documentclass[aps,prl,twocolumn,superscriptaddress]{revtex4}

\usepackage{graphicx}
\usepackage{hyperref}

\def\beq{\begin{equation}}
\def\eeq{\end{equation}}
\def\bea{\begin{eqnarray}}
\def\eea{\end{eqnarray}}
\def\bi{\begin{itemize}}
\def\ei{\end{itemize}}
\def\bp{\begin{picture}}
\def\ep{\end{picture}}
\def\bc{\begin{center}}
\def\ec{\end{center}}

\def\e0{\epsilon_0}
\def\k0{\left[\frac{1}{4\pi\e0}\right]}

\begin{document}

\title{``Light Sail'' Acceleration Revisited}
\author{Andrea Macchi}\email{macchi@df.unipi.it}
\affiliation{CNR/INFM/polyLAB, Pisa, Italy}
\affiliation{Dipartimento di Fisica ``Enrico Fermi'', Universit\`a di Pisa, Largo B. Pontecorvo 3, I-56127 Pisa, Italy}
\author{Silvia Veghini}
\author{Francesco Pegoraro}
\affiliation{Dipartimento di Fisica ``Enrico Fermi'', Universit\`a di Pisa, Largo B. Pontecorvo 3, I-56127 Pisa, Italy}

\date{\today}

\begin{abstract}
The dynamics of the acceleration of ultrathin foil targets by the 
radiation pressure of superintense, circularly polarized laser pulses 
is investigated by analytical modeling and particle-in-cell simulations.
By addressing self-induced transparency and charge separation effects, 
it is shown that for ``optimal'' values of the foil thickness 
only a thin layer at the rear side is accelerated by radiation pressure.
The simple ``Light Sail'' model gives a good estimate
of the energy per nucleon, but overstimates the conversion efficiency of laser
energy into monoenergetic ions.     
\end{abstract}

\maketitle

Radiation Pressure Acceleration (RPA) of ultrathin solid targets
by superintense laser pulses has been proposed as a promising way to 
accelerate large numbers of ions up to ``relativistic'' energies, i.e. in the 
GeV/nucleon range \cite{esirkepovPRL04,zhangPP07,robinsonNJP08,klimoPRSTAB08,yanPRL08,qiaoPRL09,gonoskovPRL09,tripathiPPCF09,rykovanovNJP08}.
The simplest model of this acceleration regime is that of a ``perfect'' 
(i.e. totally reflecting) plane mirror boosted by a light wave at perpendicular
incidence \cite{simmonsAJP92}, 
which is also known as the ``Light Sail'' (LS) model.
The LS model predicts the efficiency $\eta$, defined as the ratio between the 
mechanical energy of the mirror over the electromagnetic energy of the light 
wave pulse, to be given by
\beq
\eta={2\beta}/({1+\beta}), \qquad \beta={V}/{c},
\label{eq:eff}
\eeq
where $V$ is the mirror velocity; hence, RPA becomes more and more efficient
($\eta\rightarrow 1$) as $\beta\rightarrow 1$.
Heuristically, Eq.(\ref{eq:eff}) 
can be explained by the conservation of the number
of ``photons'' $N$ of the light wave reflected by the 
moving mirror in a small time interval:
each photon has energy $\hbar\omega$, thus the total energy of the incident 
and reflected pulses are given by $N\hbar\omega$ and $N\hbar\omega_r$,
where $\omega_r=\omega(1-\beta)/(1+\beta)$ due to the Doppler effect,
and the energy transfered to the mirror is given by their difference 
$[2\beta/(1+\beta)]N\hbar\omega$. 

The predictions of the LS model are very appealing for applications, but one
may wonder to what extent this picture is appropriate to describe the 
acceleration of a solid target by a superintense laser pulse. In the present
paper, we revisit the LS model with the help of simple modeling and 
particle-in-cell (PIC) simulations. We address issues outside the model 
itself, such as the effects of nonlinear reflectivity and charge depletion,
and on this basis we explain a few features observed in simulations. 
Our main result is that the LS model is accurate in predicting the ion energy
but overstimates the corresponding conversion efficiency, i.e. the fraction of
the laser pulse energy transferred \emph{into quasi-monoenergetic ions},
due to the fact that only a layer of the foil \emph{at its rear side}
is accelerated by RPA.

Our analysis is confined to a one-dimensional (1D) approach for the sake of 
simplicity and because multi-dimensional simulations showed that a 
``quasi-1D'' geometry has to be preserved in the acceleration stage (by using 
flat-top intensity profiles) to avoid early pulse transmission due to the
expansion of the foil in the radial direction \cite{liseikinaPPCF08}.
Circularly polarized pulses are used to reduce electron heating 
\cite{macchiPRL05}, an approach followed by several 
groups for efficient acceleration of thin foils 
\cite{zhangPP07,robinsonNJP08,klimoPRSTAB08,yanPRL08,qiaoPRL09,liseikinaPPCF08}.
We do not consider intensities high enough that ions become
relativistic within the first laser cycle; this condition may
affect the early stage of charge depletion (e.g. by narrowing the temporal 
scale separation between ions and electrons), and lead to different estimates
\cite{esirkepovPRL04,qiaoPRL09}. 

The LS model is based on the following equation of motion for the foil
\beq
\frac{d}{dt}(\beta\gamma)=\frac{2I(t-X/c)}{\rho\ell c^2}R(\omega')
                          \frac{1-\beta}{1+\beta},
\label{eq:LS}
\eeq
where $\gamma=1/\sqrt{1-\beta^2}$, $dX/dt=V$, $I$ is the light wave intensity,
$\rho$ and $\ell$ are the mass density and thickness of the foil, 
$R(\omega')$ is the reflectivity in the rest frame of the 
foil, and $\omega'=\omega\sqrt{(1-\beta)/(1+\beta)}$. 
For suitable expressions of $R(\omega')$, 
the final velocity $\beta_{\mbox{\tiny f}}$ 
can be obtained from Eq.(\ref{eq:LS}) as a function of
the pulse fluence $F=\int I dt$. For $R=1$, one obtains
\beq \label{eq:betaf}
\beta_{\mbox{\tiny f}}=\frac{(1+{\cal E})^2-1}{(1+{\cal E})^2+1},\qquad 
{\cal E}=\frac{2F}{\rho\ell c^2}
        =2\pi\frac{Z}{A}\frac{m_e}{m_p}\frac{a_0^2\tau}{\zeta}.
\eeq
In the last equality we wrote the fluence in dimensionless units as 
$a_0^2\tau$, where $a_0=\sqrt{I/m_e c^3n_c}$ is the 
dimensionless pulse amplitude and $\tau$ is the pulse duration in units of 
the laser period, and introduced the parameter 
$\zeta=\pi(n_0/n_c)(\ell/\lambda)$ which 
characterizes the optical properties of a sub--wavelength plasma foil 
\cite{vshivkovPP98}.
In these equations, $n_0$ is the initial electron density, 
$n_c=\pi m_e c^2/e^2\lambda^2$ is the cut-off density, and 
$\lambda$ is the laser wavelength. In practical units, 
$n_c=1.1 \times 10^{21}~\mbox{cm}^{-3}[\lambda/\mu\mbox{m}]^{-2}$ and 
$a_0=(0.85/\sqrt{2})(I\lambda^2/10^{18}~\mbox{W cm}^{-2}\mu\mbox{m}^2)^{1/2}$
for a circularly polarized laser pulse.
Using Eq.(\ref{eq:betaf}) it is found that
with a $1~\mbox{ps}=10^{-12}~\mbox{s}$, $1~\mbox{PW}=10^{12}~\mbox{W}$ 
laser pulse and a $10~\mbox{nm}$ target of 
$1~\mbox{g cm}^{-3}$ density, 
$\sim 1~\mbox{GeV}$ per nucleon may be obtained.
As the LS model assumes the target to be a \emph{perfect} mirror 
(i.e. rigid and totally reflecting), it implies that \emph{all} the ions 
are accelerated to the same velocity and the spectrum is 
perfectly monoenergetic.

\begin{figure}[th!]
\includegraphics[width=0.48\textwidth]{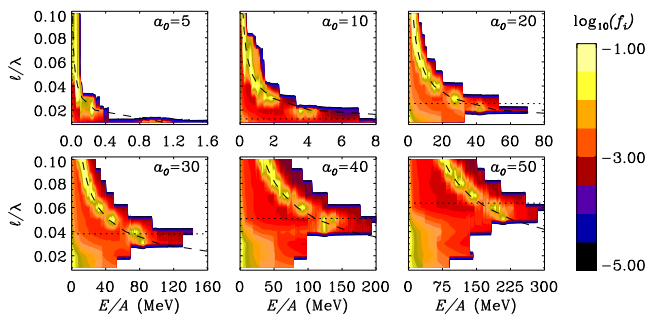}
\caption{(Color online)
Parametric study of the ion energy spectra vs. laser amplitude $a_0$ and foil
thickness $\ell$. The contours of $\log_{10}f_i(E)$ are shown, 
with $f_i(E)$ the energy per nucleon distribution normalized to unity.
For all runs, $n_0=250n_c$, $Z/A=1/2$, $\tau=9$.
The dashed line shows the prediction of the LS model for the ion energy.
The dotted horizontal line marks 
$\ell_{\mbox{\tiny opt}}$ given by the $\zeta=a_0$ condition. 
\label{fig:param}}
\end{figure}

\begin{figure}[h!]
\includegraphics[width=0.48\textwidth]{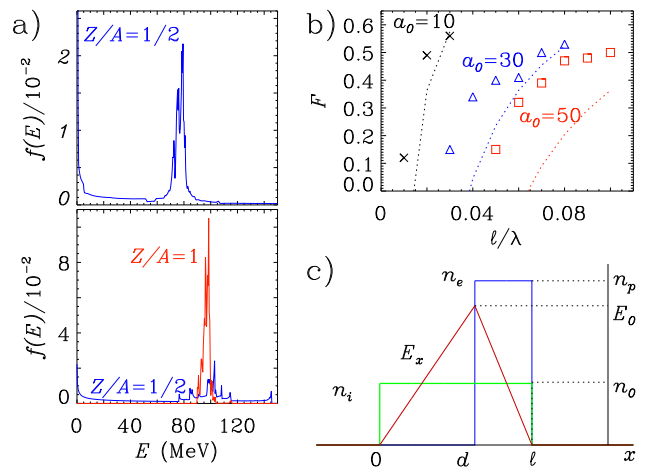}
\caption{(Color online) a): Ion energy spectra (in energy per nucleon) from a 
simulation with $a_0=30$ and a $\ell=0.04\lambda$ thick foil of a
single ion species with $Z/A=1/2$ (top) 
and one with the same parameters but where ions in a thin
surface layer ($0.01\lambda$) at the rear side are replaced by protons 
(bottom).
Fig.\ref{fig:param}.
b) Fraction of ions contained in the spectral peak 
vs. the target thickness $\ell$ for three values of 
$a_0=10$ (black, crosses), $30$ (blue, triangles) and $50$ (red, squares).
The dashed lines correspond to Eq.(\ref{eq:F}) for $F$.
All other parameters for both a) and b) are the same as in Fig.\ref{fig:param}.
c) Approximate profiles of ion ($n_i$, green) and electron ($n_e$, blue) 
densities and of the electrostatic field ($E_x$, red) in the early stage of 
the interaction, before ions move.
\label{fig:cartoon}
\label{fig:spectra}}
\end{figure}

Fig.\ref{fig:param} shows a parametric study of the ion spectrum vs. 
$\ell$ and $a_0$ from PIC simulations. For all runs, 
$n_0=250n_c$, $Z/A=1/2$ and the pulse has a flat--top envelope with 1 cycle 
rise and fall times and 8 cycles plateau. For each value of $a_0$ and for
$\ell$ less than a threshold value $\ell_{\mbox{\tiny opt}}$
we observe a narrow spectral peak, whose energy 
increases with decreasing $\ell$ and is in very good agreement with the 
predictions of the LS model, assuming $R=1$.
A typical lineout of the spectrum is shown in Fig.\ref{fig:spectra}~a).
For $\ell > \ell_{\mbox{\tiny opt}}$, the peak disappears 
and a thermal--like spectrum is observed. This is correlated with an almost
complete expulsion of the electrons from the foil in the forward direction
at the beginning of the interaction, leading to a Coulomb explosion of the 
ions. 

The results of Fig.\ref{fig:param} show that the LS model is useful for
quantitative predictions of the ion energy, but also suggest several questions
of interest both for the basic physics of RPA and its applications.
How is $\ell_{\mbox{\tiny opt}}$ determined? Does the reflectivity of the foil
and relativistic effects on the latter play a role? As the radiation pressure
tends to separate electrons from ions, does the foil remain neutral before 
and/or after the acceleration stage? Moreover, as shown shown in 
Fig.\ref{fig:spectra}~b), the ``monoenergetic'' peak contains just a fraction
of the total number of ions, and such fraction depends on $\ell$ and $a_0$.
This is different from the assumption of the LS model, which assumes all the
ions in the foil to move coherently with the foil, and may sound surprising,
since the peak energy is in agreement with the LS formula where the 
\emph{whole} mass of the foil, including low-energy ions out of spectral peak, 
is used.
In the following we provide answers to the questions above by discussing 
effects not included in the simplest LS model, i.e. beyond the description of
the foil as a perfect, rigid mirror.

First we discuss effects related to the reflectivity $R$ of the plasma
foil. For very high intensities, electrons oscillate with relativistic 
momenta in the laser field, leading to a nonlinear dependence of $R$ 
upon $a_0$. An explicit expression can be found analytically by using 
the model of a delta--like ``thin foil'' \cite{vshivkovPP98}, 
i.e. a plasma slab located at 
$x=0$ with electron density $n_e(x)=n_0\ell \delta(x)$.
The expression obtained for $R$ in the rest frame of the foil
is very well approximated by
\beq
R \simeq \left\{
\begin{array}{lr}
 {\zeta^2}/({1+\zeta^2}) & (a_0 < \sqrt{1+\zeta^2})\\
 {\zeta^2}/{a_0^2}  & (a_0 > \sqrt{1+\zeta^2})
\end{array}\right. .
\label{eq:Rapprox}
\eeq
A threshold for self-induced transparency of the foil may thus
be defined as $a_0=\sqrt{1+\zeta^2} \simeq \zeta$ when $\zeta\gg 1$, i.e.  
in most cases of interest. According to Eq.(\ref{eq:Rapprox}),
the total radiation pressure $P_{\mbox{\tiny rad}}$ on the target
\beq
P_{\mbox{\tiny rad}}=2 R{I}/{c}=2m_e c^3 n_c a_0^2 R
\eeq
becomes \emph{independent} upon $a_0$ for $a_0>\zeta$. 
Thus, the maximum radiation pressure is obtained for 
$a_0 \lesssim \zeta$, and in this condition typically $R \simeq 1$ for solid
densities. This suggests that the optimal thickness $\ell_{\mbox{\tiny opt}}$
is determined by the condition $a_0\simeq\zeta$, in good agreement with
the simulation results in Fig.\ref{fig:param} and as also found by other
studies \cite{tripathiPPCF09,esirkepovPRL06}.

The nonlinear reflectivity of the thin foil is determined by the transverse
motion of electrons (in the foil plane). However, for a thin but ``real''
target the radiation pressure tends to push electrons also in longitudinal
direction, and may remove them from the foil. 
Let us compare $P_{\mbox{\tiny rad}}$ with the electrostatic pressure 
$P_{\mbox{\tiny es}}$  that would be generated if all electrons would be 
removed from the foil.
The condition 
\beq
P_{\mbox{\tiny rad}} \geq P_{\mbox{\tiny es}}=2\pi(en_0 \ell)^2
\label{eq:balance}
\eeq
corresponds to the threshold for the removal of all electrons from the foil.
However, when Eq.(\ref{eq:Rapprox}) is used for 
$a_0 \leq \sqrt{1+\zeta^2}$, 
Eq.(\ref{eq:balance}) reduces to $a_0 \geq \sqrt{1+\zeta^2}$, while 
for $a_0 \geq \sqrt{1+\zeta^2}$ we find that 
$P_{\mbox{\tiny rad}}=P_{\mbox{\tiny es}}$ holds. It is thus possible to produce a 
density distribution where 
\emph{all electrons pile at the rear surface of the foil}. In fact, if 
$a_0 \lesssim \zeta$ and $R \simeq 1$, the laser pulse compresses the 
electron layer while keeping $R$ constant since the product $n_e\ell$ does not 
change during the compression; at the same time almost no electrons are ejected 
from the rear side because the ponderomotive force vanishes there if 
$R \simeq 1$, and so does the electrostatic field: the qualitative profiles of 
the electron density and of the electric field are shown in 
Fig.\ref{fig:cartoon}~c).    
Since for $a_0$ close to $\zeta$ the equilibrium between the electrostatic
and radiation pressures occurs only when the depth $d$ of the region of charge 
depletion is close to $\ell$, electrons are compressed in a very thin layer.
The depletion depth $d$ may be estimated from the equilibrium condition 
\begin{equation}
P_{\mbox{\tiny es}}=2\pi (e n_0 d)^2 \simeq 2I/c
\label{eq:balance2}
\end{equation}
when $R \simeq 1$, which yields $d \simeq \ell(a_0/\zeta) \lesssim \ell$. 
It is worth to point out that these considerations are appropriate for a 
circularly polarized laser pulse; for linear polarization, all electrons may be 
expelled for a transient stage under the action of the ${\bf J}\times{\bf B}$
force whose peak value per unit surface exceeds $2RI/c$ due to its 
oscillating component. Complete expulsion of electrons for $a_0>\zeta$ has
been discussed in Ref.\cite{bulanovjrPRE08}.

\begin{figure}[h!]
\includegraphics[width=0.48\textwidth]{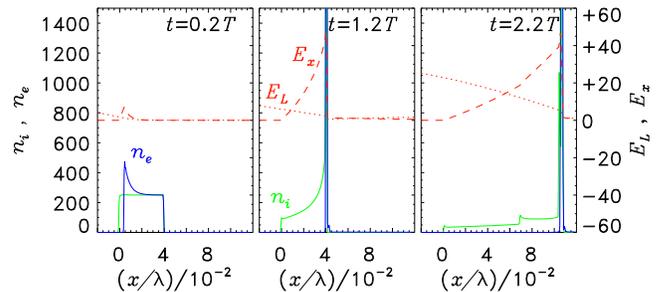}
\caption{(Color online) 
Snapshots from a 1D PIC simulation of the interaction of a 
laser pulse with a thin plasma slab. The ion density $n_i$ 
(green), the electron density $n_e$ (blue), 
the longitudinal electric field $E_x$ (red, dashed) 
and the pulse field amplitude $E_L=\sqrt{E_y^2+E_z^2}$ (red, dotted)
are shown. The target left boundary is at $x=0$
where the pulse impinges at $t=0$. Times are normalized to the
laser period $T$, fields to $m_e\omega c/e$, and densities
of $n_c$. The laser pulse has amplitude $a_0=30$ and the foil thickness
is $\ell=0.04\lambda$. All other parameters are the same of Fig.\ref{fig:param}.
\label{fig:PIC}}
\end{figure}

The snaphshots from a PIC simulation shown in Fig.\ref{fig:PIC} for a case with
$a_0=30$ and $\zeta=31.4$ confirm the scenario outlined above. The electron 
density $n_e$ reaches values (out of scale in Fig.\ref{fig:PIC}) up to tens 
of the initial density. A very high resolution 
$\Delta x=\lambda/2000$ is used to  resolve the density spike properly. 
For a laser pulse with flat--top envelope the density spiking at the rear
side of the foil is particularly evident, but we verified that it occurs also 
for a ``$\sin^2$'' envelope. Similar features were observed
also in Refs.\cite{yanPRL08,tripathiPPCF09}, but not
discussed in detail.

The electron compression in a thin layer during the initial ``hole boring'' 
stage has important consequences for the later acceleration stage.
Let us refer to the approximate field profiles in the initial stage,
sketched in Fig.\ref{fig:cartoon}~c), which were the basis of the model
presented in Ref.\cite{macchiPRL05}. This model suggests that 
\emph{only the ions located initially in the electron compression layer} 
($d<x<\ell$) \emph{will be bunched and undergo} RPA 
(via a ``cyclic'' acceleration as discussed in 
Refs.\cite{zhangPP07,robinsonNJP08,klimoPRSTAB08})
because for these ions only the 
electrostatic pressure balances the radiation pressure, while the ions 
in the electron depletion layer ($0<x<d$)
will be accelerated via Coulomb explosion, i.e. by their own 
space-charge field. 
This is exactly what is observed in the PIC simulations, both in the 
density profiles (see Fig.\ref{fig:PIC} at $t=2.2T$)
and in the ion spectra.
This effect also explains how RPA with circularly polarized 
pulses may work also in double layer targets 
\cite{esirkepovPRL06}, if the thickness of a thin layer on the
rear side matches $\ell_{\mbox{\tiny eff}}=\ell-d$.
Fig.\ref{fig:spectra} shows ion spectra for the same simulation of 
Fig.\ref{fig:PIC} and for a simulation with the same parameters, but 
where ions in a surface layer of $0.01\lambda$ thickness have been replaced by
protons. 
A fraction of heavier ions is also accelerated to the same energy per nucleon
as the protons, a typical feature of RPA of a thin plasma foil.

As an additional consequence of the piling up of electrons at the rear surface, 
the portion of the foil which is boosted by the laser pulse is negatively 
charged due to the excess of electrons. However, the simulations show that
when the laser pulse is over the excess electrons detach from the foil and
move in the backward direction, so that the accelerated layer is eventually
neutral. This is important to avoid a later Coulomb explosion of the layer and
to preserve a monoenergetic spectrum.
During the acceleration, the longitudinal field at the surface of the 
accelerated layer is almost constant implying that the charge there contained 
is also constant. It is thus possible to estimate the fraction $F$ of 
accelerated ions from the initial equilibrium condition, 
Eq.(\ref{eq:balance2}), as
\bea
F \simeq \ell_{\mbox{\tiny eff}}/\ell \simeq (1-a_0/\zeta).
\label{eq:F}
\eea
The agreement with data in Fig.\ref{fig:spectra}~b) is 
qualitative, with large deviations as $F$ becomes significantly smaller than 
one. As explained below, a lower bound on $F$ is determined by energy 
conservation.

A simple argument of force balance also explains
why the energy of the spectral peak in Fig.\ref{fig:spectra} is in
very good agreement with the predictions of the LS model where the 
\emph{initial} value $\ell$ of the foil thickness is used, while 
only a layer of thickness $\ell_{\mbox{\tiny eff}}<\ell$ is accelerated
via RPA. Let us refer again to the profiles of Fig.\ref{fig:cartoon}. 
The equilibrium condition for electrons implies
\begin{equation}
\frac{2I}{c}\doteq\int_d^{\ell}eE_0\left(\frac{\ell-x}{\ell-d}\right)n_p dx
            =\frac{1}{2}e n_0 E_0 \ell. 
\label{eq:equilb1}
\end{equation}
The electric field pushes ions in the compression layer $d<x<\ell$, 
exerting a total pressure
\begin{equation}
P_c=\int_d^{\ell}eE_0\left(\frac{\ell-x}{\ell-d}\right)n_i dx
   =\frac{2I}{c}\left(\frac{\ell-d}{\ell}\right),
\end{equation}
where we used Eq.(\ref{eq:equilb1}) and assumed $R=1$. 
The equation of motion for the ion layer,
in the early stage, can be thus written as
\begin{equation}
\frac{d}{dt}[\rho(\ell-d)\gamma\beta]=\frac{P_c}{c}
=\frac{2I}{c}\left(\frac{\ell-d}{\ell}\right) ,
\end{equation}
which is trivially equivalent to
\begin{equation}
\frac{d}{dt}(\rho\ell\gamma\beta)=\frac{2I}{c},
\end{equation}
i.e. to the equation of motion one would write for the \emph{whole} foil.
The argument may be applied also when the layer is in motion leading to 
the same conclusion. 
Having the same $\beta(t)$ as the whole foil implies that the energy per 
nucleon and the efficiency (\ref{eq:eff}) will be also the same, but the 
total kinetic energy will be lower for the thin layer. The rest of the 
absorbed energy is stored in the 
electrostatic field and as kinetic energy of the ions in the $x<X(t)$ region.
Let us consider for example  the energy stored in the electrostatic
field. At the time $t$, the field $E_x$ 
between the initial $(x=0)$ and the actual $\left(x=X(t)\right)$ 
positions of the front surface of the foil is given approximately by
$E_x=E_0 x/X(t)$, where $E_0=4\pi en_0 d$, corresponding to an 
electrostatic energy per unit surface
\beq
U_{\mbox{\tiny es}}=U_{\mbox{\tiny es}}(t)=\int_0^{X(t)}\frac{E^2_x(x,t)}{8\pi}dx,
\eeq
which varies in time as
\beq
\frac{dU_{\mbox{\tiny es}}}{dt}=\frac{1}{8\pi}E^2_x[X(t)]\frac{dX}{dt}
                  =\frac{1}{8\pi}E^2_0\beta c .
\label{eq:duesdt}
\eeq
Dividing (\ref{eq:duesdt})
by the laser intensity we obtain the ``conversion efficiency'' 
into electrostatic energy $\eta_{\mbox{\tiny es}}$
\beq
\eta_{\mbox{\tiny es}}=\frac{1}{I}\frac{dU_{\mbox{\tiny es}}}{dt}
         ={2\beta}\left(\frac{d}{\ell}\right)^2
                         \left(\frac{\zeta}{a_0}\right)^2.
\eeq
If ${\zeta}\simeq {a_0}$ and thus $d\simeq\ell$, we would obtain 
$\eta_{\mbox{\tiny es}}\simeq 2\beta > \eta$ that is unphysical. 
Thus, the energy stored in the
electrostatic field also prevents the accelerated layer thickness to 
shrink to zero. 

In conclusion, we have revisited the ``Light Sail'' model of Radiation
Pressure Acceleration of a thin plasma foil. The nonlinear reflectivity of the 
foil determines the ``optimal'' condition $\zeta \simeq a_0$, for which the 
energy in the RPA spectral peak is highest and in good agreement with the
LS model formula where the \emph{total} thickness (or the total mass) of the 
foil enters as a parameter. However, not all the foil is accelerated, but only 
a thin layer at the rear side of thickness $\ell_{\mbox{\tiny eff}}<\ell$; 
the apparent paradox is solved by observing
that, to keep electrons in a mechanical quasi-equilibrium, the electrostatic 
pressure pushing ions in the accelerated layer is 
$\ell_{\mbox{\tiny eff}}/\ell$ times the radiation pressure on electrons,
so that the equation of motion for the thin layer is the same as if the
whole foil were accelerated. Finally, we showed that the energy stored in
the electrostatic field is comparable to the kinetic energy and must be
taken into account. For applications, the most
relevant consequences and differences with respect to the simplest LS picture  
are that the number of ``monoenergetic'' ions is reduced, so that the 
actual efficiency may be quite lower than given by Eq.(\ref{eq:eff}), and that 
also light ions in a thin layer at the rear surface (e.g., hydrogen impurities) 
may be accelerated by RPA.

Support from CNR via a RSTL project and use of supercomputing facilities at
CINECA (Bologna, Italy) sponsored by the CNR/INFM supercomputing initiative 
are acknowledged.

\bibliographystyle{apsrev}

\hyphenation{Post-Script Sprin-ger}

\end{document}